# Water-induced superconductivity in SrFe$_2$As$_2$

**Hidenori Hiramatsu** [1, *], **Takayoshi Katase** [2], **Toshio Kamiya** [1, 2], **Masahiro Hirano** [1, 3], **and Hideo Hosono** [1, 2, 3]

1: ERATO–SORST, Japan Science and Technology Agency, in Frontier Research Center, Tokyo Institute of Technology, S2-6F East, Mailbox S2-13, 4259 Nagatsuta-cho, Midori-ku, Yokohama 226-8503, Japan

2: Materials and Structures Laboratory, Mailbox R3-1, Tokyo Institute of Technology, 4259 Nagatsuta-cho, Midori-ku, Yokohama 226-8503, Japan

3: Frontier Research Center, S2-6F East, Mailbox S2-13, Tokyo Institute of Technology, 4259 Nagatsuta-cho, Midori-ku, Yokohama 226-8503, Japan

**ABSTRACT**

It has been considered that FeAs-based high transition temperature (high-$T_c$) superconductors need electron or hole doping by aliovalent ion substitution or large off-stoichiometry in order to induce superconductivity. We report that exposure of undoped SrFe$_2$As$_2$ epitaxial thin films to water vapor induces a superconducting transition. These films exhibit a higher onset-$T_c$ (25 K) and larger magnetic field anisotropy than those of cobalt-doped SrFe$_2$As$_2$ epitaxial films, suggesting that the mechanism for the observed superconducting transition differs from that of the aliovalent-ion doped SrFe$_2$As$_2$. The present finding provides a new approach to induce superconductivity with a higher $T_c$ in FeAs-based superconductors.

(*) Electronic mail: h-hirama@lucid.msl.titech.ac.jp

Discovery of an Fe-based layered superconductor, F-doped LaFeAsO rekindled extensive researches on superconducting materials [1]. Since then, four types of materials systems have been known to be superconductors, which are $Re$FeAsO ($Re$ = rare earths) [1-3], $Ae$Fe$_2$As$_2$ ($Ae$ = alkaline earths) [4], $A$FeAs ($A$ = alkali metal) [5] and FeSe [6] phases, and the maximum transition temperature ($T_c$) has reached 56 K in Th-doped GdFeAsO [3]. Each of the Fe-based compounds has layered crystal structures that possess FeAs (or FeSe) layers composed of edge-sharing FeAs$_4$ (FeSe$_4$) tetrahedra, implying that the two-dimensionality and the local FeAs (FeSe) structures play an important role in inducing the superconductivity. Mostly, undoped parent phases of these materials exhibit structural phase transitions, which accompany antiferromagnetic ordering transitions, at low temperatures, but do not exhibit superconductivity at least down to 2 K [6, 7-10]. Carrier doping by aliovalent ion substitution [1-4, 11-13] or introducing off-stoichiometry [5, 6, 14, 15] suppresses the structure and magnetic transitions and induces superconductivity. In addition to the carrier doping, high-pressure-induced superconductivity have been reported in $Ae$Fe$_2$As$_2$ [16-19]. Further, there has been one report on ambient-pressure superconductivity on nominally undoped $Ae$Fe$_2$As$_2$ [20]; however, it is considered that the superconductivity in the nominally-undoped cases may be caused by unintentional carrier doping, but the details have not been clarified yet. These results indicate that superconductivity in $Ae$Fe$_2$As$_2$ is sensitive to an external stress, a sample preparation condition, and a post-treatment condition; it in turn suggests that we may find a new approach to attain high-$T_c$ superconductivity in the FeAs-based compounds.

In this letter, we report a new type of a superconducting transition in undoped SrFe$_2$As$_2$, in which exposure of SrFe$_2$As$_2$ epitaxial thin films to water vapor invokes a superconducting transition at an onset transition temperature ($T_c^{onset}$) of 25 K. It would be worth noting that water-intercalated superconductivity is known in some layered crystals such as $A_x$(H$_2$O)$_y$$T$S ($A$ = Na, K, Rb, and Cs; $T$ = Nb, and Ta; $T_c$ = 5.5 K) [21], Na$_x$CoO$_2$·$y$H$_2$O ($T_c$ = 5 K) [22], and Na$_x$(H$_2$O)$_y$($M$S)$_{1+\delta}$(TaS$_2$)$_2$ ($M$ = Sn, Pb, and Sb; $T_c$ = 4 K) [23]. These are explained by enhanced two-dimensionality in connection with the expansion of the inter-layer distances induced by the water intercalation. On the contrary, the SrFe$_2$As$_2$ epitaxial films exhibit the superconducting transition although the inter-layer distance (i.e. the $c$-axis length of the unit cell) is shrunken by the exposure to water vapor. In addition, the $T_c$ and the superconductivity anisotropy are different largely from aliovalent-ion doped $SrFe_2As_2$. These discrepancies imply that a new mechanism controls the emergence of the superconducting transition in $SrFe_2As_2$ by water exposure.

Epitaxial thin films of undoped $SrFe_2As_2$ (thickness ~ 200 nm) were grown on mixed perovskite (La, Sr)(Al, Ta)$O_3$ (LSAT) (001) single-crystal substrates by pulsed laser deposition [24, 25] using undoped $SrFe_2As_2$ sintered disks as the targets at the substrate temperature of ~700 °C in vacuum of ~$10^{-5}$ Pa. Crystalline quality and orientation were examined by high-resolution x-ray diffraction (XRD, radiation: CuK$\alpha_1$, apparatus: ATX-G, Rigaku). We confirmed that the intensity and width of diffraction peaks and rocking curves, which are the measure of the crystalline quality and orientation, were comparable to those of the epitaxial films of cobalt-doped $SrFe_2As_2$ that exhibited superconducting transitions at $T_c^{onset}$ = 20 K [25]. Resistivity ($\rho$) – temperature ($T$) curves were measured using a physical properties measurement system (PPMS, Quantum Design) in vacuum from 2 to 300 K under external magnetic fields ($H$) varied up to 9 T.

The solid line in Fig. 1(a) shows the $\rho-T$ curve of the virgin $SrFe_2As_2$ epitaxial film, which was measured immediately after taken out from the deposition chamber. A resistivity anomaly corresponding to a structural phase transition and an antiferromagnetic ordering is observed at $T_{anom}$ = 204 K, which is almost the same as those reported previously for undoped bulk samples [26, 27]. No superconducting transition is observed for the virgin film. On the other hand, after exposing the virgin film to an ambient atmosphere at room temperature with the relative humidities of 40 – 70 RH% for 2 hours, $\rho$ started dropping at 25 K [the dashed line in Fig. 1(a) and the triangles in Fig. 1(b)]. With increasing the exposure time, the drop of $\rho$ became sharper, and finally the zero resistance was observed at the exposure time ≥ 4 hours. The $T_c^{onset}$ and the offset $T_c$ of the film exposed to air for 6 hours (hereafter, referred to as the air-exposed film) are 25 K and 21 K, respectively, which are higher by ~5 K than those of cobalt-doped $SrFe_2As_2$ epitaxial films [25]. These observations strongly suggest that a constituent in an ambient air induces the superconducting transition in $SrFe_2As_2$.

Figures 1(c) and (d) show magnetic anisotropy of the air-exposed film. $T_c$ was shifted to lower temperatures with increasing $H$ for both the $H$ directions parallel to the $c$-axis ($H_{//c}$) and the $a$-axis ($H_{//a}$).

Superconducting transitions survived at ≤ 7 K for $H_{//c}$ and ≤ 16 K for $H_{//a}$ even when $H$ was raised up to 9 T, indicating that the upper critical magnetic fields are far above 9 T. It is observed that $T_c$ is more sensitive to the $H_{//c}$ than to the $H_{//a}$. It would be worth noting that the large anisotropy of $T_c$ in the air-exposed $SrFe_2As_2$ epitaxial films is rather different from that of the cobalt-doped $SrFe_2As_2$ epitaxial films reported in refs. [25, 28], which showed almost isotropic behaviors; the large anisotropy is more similar to that reported for the $Re$FeAsO system [29].

Figure 2(a) shows out-of-plane XRD patterns around the $SrFe_2As_2$ 002 diffraction. It shows that exposure to air for 6 hours broadened and upper-shifted the $SrFe_2As_2$ 002 diffraction peak. It accompanies formation of a small amount of an $Fe_2As$ impurity phase, which is confirmed by an additional broad peak at $2\theta \sim 14.9$ degrees. This observation indicates that the $c$-axis length of the $SrFe_2As_2$ phase was decreased from 1.233 nm in the virgin film (this value is slightly smaller (–0.4%) than that of bulk $SrFe_2As_2$, 1.238 nm [26]) to 1.227 nm in the air-exposed film. The latter value is smaller even than that of the cobalt-doped $SrFe_2As_2$ epitaxial films ($c$ = 1.230 nm [25]). This shrinkage would be related to the superconducting transition because it has been reported that applying a high pressure to undoped $SrFe_2As_2$ invokes superconducting transitions [17-19]. In those cases, $T_c^{onset}$ reaches ca. 35 – 38 K and $T_{anom}$ shifts to lower temperatures as the external pressure increases [18, 19]. On the other hand, in the present case, the shift of $T_{anom}$ is not observed as seen in Fig. 1(a), and the observed $T_c$ are lower than the maximum $T_c$ of the pressure-induced cases. Fig. 2(b) shows the Williamson-Hall plots of the full width at half maximum (FWHM, $\Delta$) for the 00$l$ diffractions of the $SrFe_2As_2$ films, which corresponds to the relation $\Delta \cdot \cos\theta / \lambda = K (D^{-1} + 2\eta \cdot \sin\theta / \lambda)$ where $\theta$ denotes the diffraction angle, $K$ the constant (0.9 for a Gaussian peak profile), $D$ the height of the single-domain crystallite, $\lambda$ the wavelength of x-ray, and $\eta$ the inhomogeneous strain [30]. The slope ($2\eta$) was not changed largely by the air exposure, indicating the major origin of the peak broadening is not the internal inhomogeneity in the crystalline lattice, which is often induced by lattice strain and impurity incorporation. The intercepts at the vertical axis in Fig. 2(b) provide the crystallite heights $D$, which were estimated to be 680 nm for the virgin film and 80 nm for the air-exposed film. The former value overestimates the crystallite height because the film thickness is thinner, 190 nm. This value is, however, reasonable because the intersect value approaches zero for a larger crystallite size in the Williamson-Hall plot, which gives a larger error. Therefore, this result indicates that the crystallites in the virgin film are single-domain in height. The value of the crystallite height for the air-exposed film is more accurate because the intersect value is much larger than zero, and the obtained crystallite size (80 nm) is consistent with the film thickness (190 nm). It safely concludes the air-exposure decreased the crystallite height, which is the major origin of the broadening in the diffraction peak discussed above. It is possible to estimate the change of the volume of the $SrFe_2As_2$ phase by comparing integrated diffraction intensities of the $SrFe_2As_2$ 002 diffractions, which were estimated from the intensities and widths of the diffraction peaks and the rocking curve peaks. We observed that the air-exposure reduced the peak intensity to ~20 %, and increased the width of the diffraction peak to 190 % and that of the rocking curve to 120 % of those of the virgin film, which indicates that the volume of the $SrFe_2As_2$ phase in the air-exposed film is decreased to 46 % of the virgin film. This volume change corresponds reasonably to the ratio of the residual crystallite height (80 nm) and the film thickness (190 nm), suggesting that the vertical single-domain structure is maintained in the $SrFe_2As_2$ phase even after the air exposure. By contrast, the volume fraction of the $Fe_2As$ phase was estimated to be ~10 % of the $SrFe_2As_2$ phase from the integrated peak area and the structure factors of these diffractions. These results substantiate that the $SrFe_2As_2$ is the majority phase even after the air-exposure.

To determine the origin of the appearance of the superconducting transition induced by the air exposure, we investigated the effects of exposure to the air constituents separately (Fig. 3). Exposure to a dry nitrogen gas (purity: 6N, dew point: ≤ –80 °C, pressure: 300 Torr) for 24 hours at room temperature did not induce any change [Fig. 3(a)], proving that dry $N_2$ is inert for undoped $SrF_2As_2$ films at least at room temperature. Exposure to a dry oxygen gas (6N, ≤ –80 °C, 300 Torr) and a dry carbon dioxide gas (4N, ≤ –76 °C, 760 Torr) for 24 hours at room temperature [Figs. 3(b) and (c)] gave small changes in the $\rho$ values, but did not induce a superconducting transition at least down to 2 K. The small increase in $\rho$ in the dry oxygen gas and the small decrease in the dry carbon dioxide gas suggest some doping effects by these molecules. Figure 3(d) shows the effect of water vapor (a dew point +13 °C, 760 Torr, generated from distilled water with a dry nitrogen gas as a carrier gas). A clear superconducting transition was observed at $T_c^{onset}$ = 25 K, which is the same temperature as that of the air-exposed film. Therefore, we conclude that the superconducting transition is induced by incorporation of $H_2O$-related species to the $SrFe_2As_2$ films. As clarified by the XRD study, the incorporation of water-related species accompanies the shrinkage of the c-axis length of the $SrFe_2As_2$ lattice and the appearance of the $Fe_2As$ phase. The effect of the $Fe_2As$ phase is excluded from the origin of the superconducting transition, because no superconducting transition has been reported for $Fe_2As$ (reported to be an anti-ferromagnet with the Néel temperature of 353 K [31]) as far as we know. Therefore, we consider that the shrinkage of the c-axis would closely be related to the appearance of the superconducting transition.

The crystal structure [27, 32] in Fig. 4 tells that it has two large interstitial sites; one is the $I_9$ site surrounded by four As, four Fe and one Sr [Fig. 4(b)], and the other the $I_6$ site surrounded by four Sr and two As [Fig. 4(c)], in which the sizes of the free spaces (0.15 nm in the a-b planes) are close to that of an $O^{2-}$ ion (0.14 nm). Preliminarily first-principles structure relaxation calculations [33] showed that incorporation of an OH or an $H_2O$ molecule at the $I_9$ or $I_6$ sites formed quantum-mechanically (meta-) stable structures, but it moved the molecules from the interstitial sites to inter-layer sites and expanded the c-axis length significantly, which is similar to the intercalation cases such as $A_x(H_2O)_yTS$, $Na_xCoO_2$ $yH_2O$ and $Na_x(H_2O)_y(MS)_{1+\delta}(TaS_2)$ [21-23]; it is inconsistent with the observed shrinkage of the c-axis length observed above. Incorporation of an oxygen atom at the $I_9$ site or the $I_6$ site found the stable positions at the symmetric positions near the centers of the coordination polyhedra. The former did not expand the c-axis largely (the expansion in the c-axis length is as small as 0.03 nm), while the latter expanded it by 0.12 nm because the interatomic distance between the two As atoms are only 0.35 nm and is not large enough to incorporate an oxygen ion. These results suggest that the most probable case among them is the incorporation of oxygen atoms at the $I_9$ sites, but any of these cases appears not to be able to explain the shrinkage of the c-axis length. Another possibility would be the formation of Sr vacancies by removing the Sr ions through a reaction e.g. $SrFe_2As_2 + 2H_2O$ –> $SrFe_2As_2$:$V_{Sr}$ + $Sr(OH)_2$ + $H_2$, which likely occurs because an $Ae$ element easily reacts with $H_2O$ to form a hydroxide.

In summary, we have found that water vapor induces a superconducting transition in undoped $SrFe_2As_2$ epitaxial films at an ambient pressure. The $T_c^{\text{onset}}$ was 25 K, which is higher than that of the cobalt-doped epitaxial films. It is concluded that the superconducting transition originates from the $SrFe_2As_2$ phase because $SrFe_2As_2$ is the major phase in the air-exposed film, $T_c$ is higher but similar to those reported on cobalt-doped $SrFe_2As_2$, and no other superconducting phase was detected by the XRD measurements. We can, however, not exclude possibility that a sub-product of the reaction between the $SrFe_2As_2$ film and $H_2O$ causes the superconducting transitions. Even though, such a case will lead to more interesting discovery because the remaining possible origins are $Fe_2As$, $FeAs$, and an amorphous phase that was not detected by XRD measurements, but any of them has not been reported to be a superconductor. Although further study is needed to clarify the origin of the superconduction and the doping mechanism, the present finding provides not only a new insight into understanding superconducting transitions in undoped $Ae\text{Fe}_2As_2$, but also a clue to obtaining higher $T_c$ by new doping method such as chemical modification.

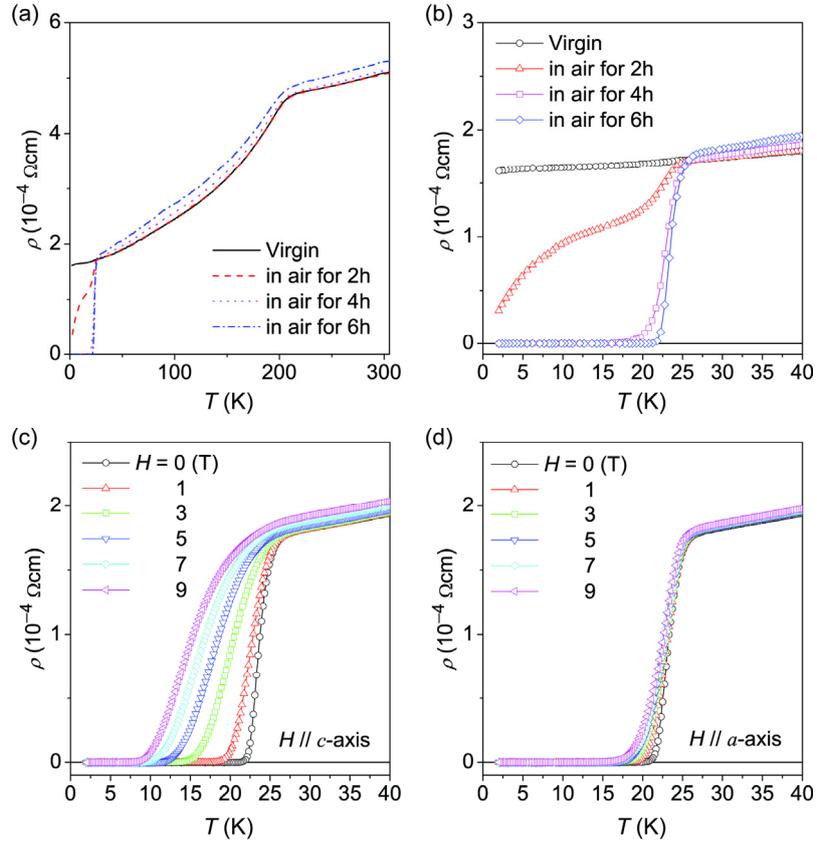

FIG. 1 (Color online). (a) Variation of $\rho-T$ curves as a function of air-exposure time for of $SrFe_2As_2$ epitaxial films as-grown (virgin, solid line), exposed to air for 2 hours (dashed line), 4 hours (dotted line), and 6 hours (dotted-dashed line). (b) Extended view of Fig. (a) between 2 and 40 K. (c, d) Anisotropy of superconducting transition of the film exposed to air for 6 hours under magnetic fields parallel to the $c$-axis ($H_{//c}$) (c) and $a$-axis ($H_{//a}$) (d) at $H_{//c,a}$ from 0 to 9 T.

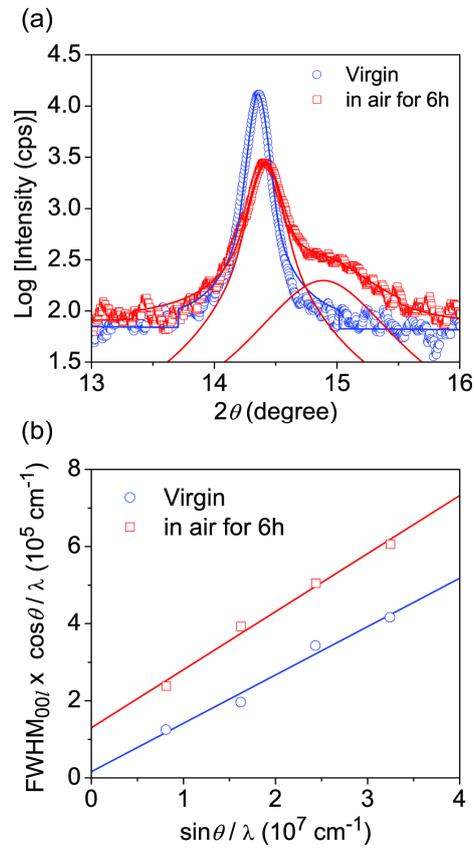

FIG. 2 (Color online). (a) $\omega$-coupled $2\theta$ scan XRD patterns around 002 diffraction of SrFe$_2$As$_2$ epitaxial films at room temperature. The circles and squares show the virgin film and the film exposed to air for 6 hours, respectively. The lines in (a) are the deconvolution of the XRD patterns. (b) Williamson-Hall plots of 00$l$ diffractions.

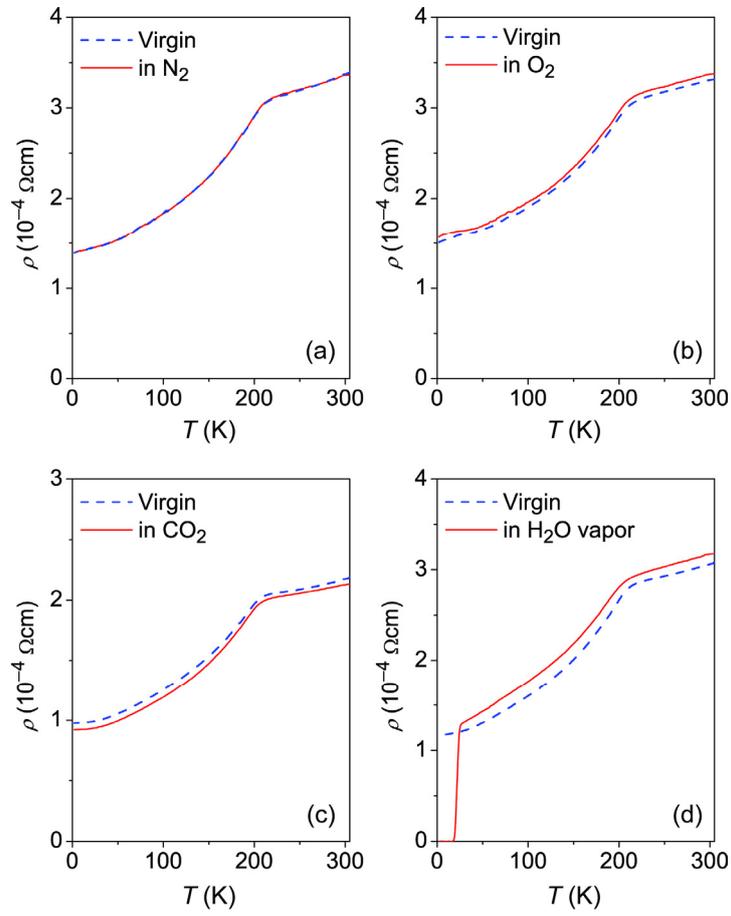

FIG. 3 (Color online). Changes in $\rho - T$ curves of $SrFe_2As_2$ epitaxial films under exposure to various atmospheres. The dashed lines show the data of the virgin films, and the solid lines those of the films exposed to dry nitrogen for 24 hours (a), dry oxygen for 24 hours (b), dry carbon dioxide for 24 hours (c), and water vapor (the dew point = + 13 °C) for 2 hours (d).

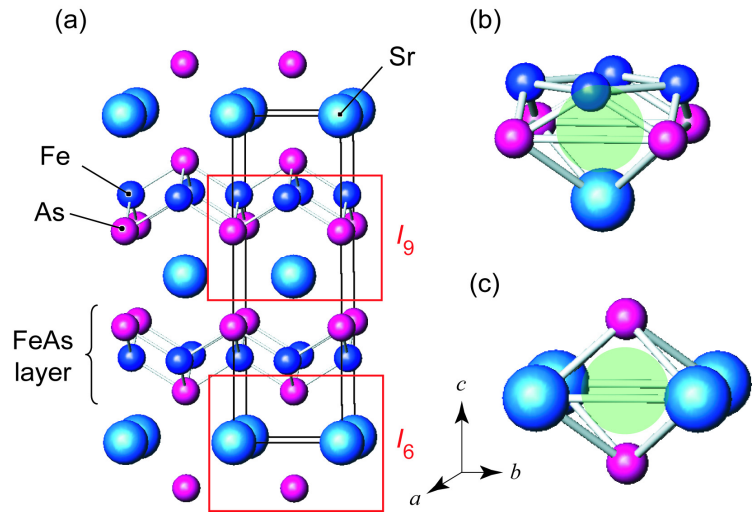

FIG. 4 (Color online). (a) Crystal structure of SrFe$_2$As$_2$ with ThCr$_2$Si$_2$-type structure. The box indicates the unit cell. (b, c) Expanded views of the interstitial sites (b) $I_9$, which is surrounded by four As, four Fe and one Sr, and (c) $I_6$, which is surrounded by four Sr and two As.